\documentclass[12pt]{amsart}
\usepackage[english]{babel}
\usepackage{a4wide,color,graphicx,amsmath,amssymb,verbatim,mathrsfs,latexsym}

\newtheorem{theorem}{Theorem}

\title[Fluid-dynamic models for graphene]{Some fluid-dynamic models for quantum electron transport in graphene via entropy minimization}
\author{Nicola Zamponi}
\address{Dipartimento di matematica Ulisse Dini, Viale Morgagni 67/A, Firenze, Italy.}
\email{nicola.zamponi@math.unifi.it}

\subjclass{Primary: 35Q40, 76Y05; Secondary: 82D37.}
\keywords{Quantum hydrodynamics, electron transport, graphene, quantum entropy principle.}

\begin{document}
%
%

\begin{abstract}
We derive some fluid-dynamic models for electron transport near a
Dirac point in graphene. We start from a kinetic model constituted
by a set of spinorial Wigner equations, we make suitable scalings
(hydrodynamic or diffusive) of the model and we build moment
equations, which we close through a minimum entropy principle. In
order to do this we make some assumptions: the usual
semiclassical approximation ($\hbar\ll 1$), and two further
hypothesis, namely
Low Scaled Fermi Speed (LSFS) and Strongly Mixed State (SMS),
which allow us to explicitly compute the closure.
\end{abstract}

\maketitle

\section{Introduction}

\subsection{Graphene electronic properties}
Graphene is a single layer of carbon atoms disposed as an honeycomb
lattice, that is, a single sheet of graphite. This new material has
attracted the attention of many physicists and engineers thanks to its
remarkable electronic properties, which make it an ideal candidate for the
construction of new electronic devices with strongly increased
performances with respect to the usual silicon semiconductors
 \cite{ AuerSchurrerErtler, Beenakker, Freitag08, KatsnelsonEtAl06}.
The great interest around graphene is attested
 by the Nobel prize gained in 2010 by Geim and Novoselov for its discovery.

Physically speaking, graphene is a zero-gap semiconductor: in the energy spectrum
the valence band intersects the conduction band in some isolated points,
named {\em Dirac points};
moreover, around such points the electron energy is approximately linear with respect to the modulus of momentum.
The system Hamiltonian can be approximated, for low energies and in
absence of potential, by the following Dirac-like operator\footnote{
We recall the Pauli matrices
$$\sigma_0 = \begin{pmatrix} 1 & 0\\ 0 & 1\end{pmatrix}, \quad
\sigma_1 = \begin{pmatrix}  0 & 1\\ 1 & 0 \end{pmatrix}, \quad
\sigma_2 = \begin{pmatrix}  0 & -i\\ i & 0 \end{pmatrix}, \quad
\sigma_3 = \begin{pmatrix}  1 & 0\\ 0 & -1\end{pmatrix},$$
where, for later convenience, we added the identity matrix $\sigma_0$.}:	
\begin{equation}\label{H0}
H_0 = -i\hbar v_F \left(\sigma_1\frac{\partial}{\partial r_1}
 + \sigma_2\frac{\partial}{\partial r_2}\right),
\end{equation}
where $v_F \approx 10^6\, \textrm{m/s}$ is the Fermi speed (at zero temperature) and, as usual,
$\hbar$ denotes the reduced Planck constant; however, in this paper we
are not going to use \eqref{H0} as the system Hamiltonian, because a
rigorous discussion of a fluid-dynamic model involving \eqref{H0}
would require considering the Fermi-Dirac entropy instead of the
Maxwell-Boltzmann, due to the lower unboundedness of the energy spectrum of
\eqref{H0} (we defer such a discussion to a future work). Indeed,
in the rest of the paper we will make the hypothesis that
the system Hamiltonian is well approximated by the following
functional:\footnote{The Weyl quantization of a symbol $\gamma$ is the
  functional $\textnormal{Op}_\hbar(\gamma)$ such that, for all $\psi$ test functions,
$$(\textnormal{Op}_\hbar(\gamma)\psi)(x) = (2\pi\hbar)^{-2}\int_{\mathbb{R}^2\times\mathbb{R}^2}
\gamma\left(\frac{x+y}{2},p\right)\psi(y)e^{i(x-y)\cdot p/\hbar}\,dydp
$$
(for more details see Ref.~\cite{Folland89}).}
\begin{equation}\label{H}
H = \textnormal{Op}_\hbar\bigg(\frac{|p|^2}{2m}\sigma_0 + v_F\vec{\sigma}\cdot\vec{p}\bigg) =
H_0 - \left(\frac{\hbar^2}{2m}\Delta\right)\sigma_0\,,
\end{equation}
with $m>0$ parameter (with the dimensions of a mass), whose
energy spectrum is bounded from below.

\subsection{Quantum fluid-dynamic models, generalities}
Recently, mathematical models of fluid-dynamic type has been developed
in order to describe quantum transport in semiconductors
\cite{JSP10, DegondRinghofer02, DegondRinghofer03, DegondMehatsRingh05,
JungelMatthesMilisic06, Jungel09, Jungel10,   Mehats09,
ZamponiBarletti}.
Such models rely on a kinetic
formulation of quantum mechanics (QM) by means of Wigner-type
equations, and are built by taking suitable moments of these latter;
the resulting equations involve the chosen fluid-dynamic moments and
usually additional expressions (referred to as \emph{not closed} terms)
which cannot be
written as functions of the previous moments without further
hypothesis. In order to solve the so-called \emph{closure problem},
that is, to compute the not closed terms from the known moments, many
techniques are employed, e.g. the pure-state hypothesis (which allows
to obtain, for a scalar Hamiltonian of type $\hat{H} =
-\frac{\hbar^2}{2m}\Delta + V(x)$, the so-called \emph{Madelung
  equations}, see \cite{Mad27}), the ad-hoc ansatz (like the Gardner's
equilibrium distribution, see \cite{Jungel10}), and a strategy of entropy
minimization (which will be followed in this paper, in analogy to the
method employed in the closure of classical fluid-dynamic systems
derived from the Boltzmann transport equation in the classical
statistical mechanics, see \cite{DegondRinghofer02, DegondRinghofer03,
DegondMehatsRingh05,  Levermore}).

The main advantages of fluid-dynamic models with respect to ''basic'' tools like Schr\"odinger,
Von Neumann, Wigner equations, are basically two.
The first advantage is about physical interpretation:
fluid-dynamic models contain already the most physically interesting
quantities (like particle, momentum and spin densities), while other
models usually involve more ''abstract'' objects (such as wavefunctions,
density operators, Wigner functions), which do not have an immediate
physical interpretation; in this latter case, further computations
have to be made in order to obtain the quantities of real physical
interest from the solution of the model.
The second and most important advantage is about
numerical computation: fluid-dynamic models for quantum systems
with $d$ degrees of freedom are sets of PDEs in $d$ space variables
and 1 time variable, while other models usually have more complicated
structures (for example, Wigner equations are sets of PDEs in $2d$
space variables and 1 time variable); so fluid-dynamic models are
usually more easily and quickly solvable via numerical computation
than other models.

In the following we will present two types of models for quantum
electron transport in graphene:
\begin{itemize}
\item Quantum Diffusion Equations (QDE);
\item Quantum Hydrodynamic Equations (QHE).
\end{itemize}
Both classes of models are obtained through the following strategy.
First, we consider the Wigner equations for the system and we perform on
them a diffusive or hydrodynamic scaling; then we take moments of
the scaled Wigner system, so that we obtain suitable sets of fluid-dynamic
equations containing not closed terms; finally we close these latter by choosing
the Wigner function equal to an equilibrium distribution,
defined as a minimizer of a suitable
entropy functional under the constraints of given moments.
We will consider here a kinetic model for quantum transport in
graphene, derived from the one-particle Hamiltonian \eqref{H},
and we will build four (semiclassical) fluid-dynamic
models for the
system under consideration: two QDE models and two QHE models.
We will perform different hypothesis and
approximations during the derivation of the fluid-dynamic models in
order to overcome the not trivial computational difficulties arising
from the process, putting in evidence (when possible) their physical
meaning.

The paper is organized as follows. In section 2 we present the kinetic
model for quantum transport in graphene that we are going to employ
in the paper. In section 3 we build a first diffusive model; in
section 4 we build a second
diffusive model; in section
5 we build a first hydrodynamic model;
in section 6 we build a second hydrodynamic model;
in section 7 we present our
conclusions and intentions for future work.

\section{A kinetic model for graphene}
Let $w = w(r,p,t)$ the system Wigner function, defined for
$(r,p,t)\in\mathbb{R}^2\times\mathbb{R}^2\times(0,\infty)$. Notice that, due to the
presence of the spin, $w$ is a complex hermitian matrix-valued
function instead of a real scalar function; so we can write $w =
\sum_{s=0}^3 w_s\sigma_s$ with $w_s$
Pauli components of $w$.
\footnote{
Given a complex hermitian $2\times 2$ matrix $a$, its Pauli components are real
numbers given by:
$$a_s = \frac{1}{2}\textnormal{tr}(a\sigma_s)\qquad s = 0,1,2,3\,.$$
}
Moreover let:
$$\vec{w} = (w_1, w_2, w_3)\,,\,\,\,\partial_t = \frac{\partial}{\partial t}\,,\,\,\,
\vec{\nabla} = \left(
\frac{\partial}{\partial_{r_1}},\frac{\partial}{\partial_{r_2}}, 0
\right)\,,\,\,\,\vec{p} = (p_1, p_2,
0)\,,\,\,\, p = (p_1,p_2)\,.$$
The Wigner equations for quantum transport in graphene, associated
with the one-particle Hamiltonian $H + V$, with $H$ defined by \eqref{H}, are:
\begin{equation}\label{WE2}
\begin{split}
\partial_t w_0 + \left[\frac{\vec{p}}{m}\cdot\vec{\nabla}\right]w_0 + v_F \vec{\nabla}\cdot\vec{w} + \Theta_\hbar(V)w_0 =&
\frac{g_0 - w_0}{\tau_c}
\\
\partial_{t}\vec{w} + \left[\frac{\vec{p}}{m}\cdot\vec{\nabla}\right]\vec{w} + v_F\Big[ \vec{\nabla} w_0 + \frac{2}{\hbar} \vec{w}\wedge\vec{p}\Big] + \Theta_\hbar(V)\vec{w} =&
\frac{\vec{g} - \vec{w}}{\tau_c}
\end{split}
\end{equation}
with the pseudo-differential operator $(\Theta_\hbar(V)w)(r,p)$ is defined by:
\begin{align*}
(\Theta_\hbar(V)w)(r,p) =& \frac{i}{\hbar}(2\pi)^{-2}\int_{\mathbb{R}^2\times\mathbb{R}^2} \delta V(r,\xi)w(r,p')
e^{-i(p-p')\cdot\xi} d\xi dp'\,,\\
\delta V(r,\xi) =& V\bigg(r + \frac{\hbar}{2}\xi\bigg) - V\bigg(r - \frac{\hbar}{2}\xi\bigg)\,.
\end{align*}
We refer to \cite{Zamponi_tesi, ZamponiBarletti} for details about the derivation of
\eqref{WE2} from the Von Neumann equation. The terms on the right side
of \eqref{WE2} are relaxation terms of BGK type, with $g$ the local
thermal equilibrium Wigner distribution, which will be defined later,
and $\tau_c$ is the mean free time (the mean time interval between two
subsequent collisions experienced by a particle).

\section{A first diffusive model for graphene.}
In the next two sections we will build two diffusive models
for quantum electron
transport in graphene starting from the kinetic model
\eqref{WE2}.
\subsection{Wigner equations in diffusive scaling}
In order to construct diffusive models for graphene,
we make the following diffusive scaling
of the Wigner equations:
\begin{equation}\label{DSCAL-1}
r\mapsto \hat{r} r\,,\quad t\mapsto \hat{t} t\,,\quad p\mapsto \hat{p}
p\,,\quad V\mapsto \hat{V} V\,,
\end{equation}
with $\hat{r}$, $\hat{t}$, $\hat{p}$, $\hat{V}$ satisfying:
\begin{equation}\label{DSCAL-2}
\frac{2 v_F \hat{p}}{\hbar} = \frac{\hat{V}}{\hat{r} \hat{p}}\,,\quad
\frac{2 \hat{p} v_F \tau_c}{\hbar} = \frac{\hbar}{2 \hat{p} v_F \hat{t}}\,,\quad
\hat{p} = \sqrt{m k_B T}\,;
\end{equation}
moreover let us define the \emph{semiclassical parameter} $\epsilon$, the
\emph{diffusive parameter} $\tau$ and the \emph{scaled Fermi speed} $c$ as:
\begin{equation}\label{DSCAL-3}
\epsilon = \frac{\hbar}{\hat{r} \hat{p}}\,,\quad
\tau = \frac{2\hat{p}v_F\tau_c}{\hbar}\,,\quad
c = \sqrt{\frac{m v_F^2}{k_B T}}\,.
\end{equation}
Notice that, if we choose as $m$ the electron mass $m_e$, then
$c^2 = \mathcal{E}_F/\mathcal{E}_{cl}$ is the ratio between the Fermi
energy $\mathcal{E}_F = m_e v_F^2$ and the classical energy $\mathcal{E}_{cl}  = k_B T$
of the electrons.

We perform two main approximations here: the well-known
\emph{semiclassical hypothesis} $\epsilon\ll 1$, and the following
assumption, which we call \emph{Low Scaled Fermi Speed} (LSFS):
\begin{equation}\label{LSFS}
c = O(\epsilon)\,.
\end{equation}

By performing the scaling \eqref{DSCAL-1}--\eqref{DSCAL-3} on the
equations \eqref{WE2} under the previous hypothesis,
 we obtain the following scaled Wigner system:
\begin{equation}\label{WDIFF2}
\begin{split}
\tau\partial_t w_0 + T_0(w) =& \frac{g_0[n] - w_0}{\tau}\,,\\
\tau\partial_t w_s + T_s(w) =& \frac{g_s[n] - w_s}{\tau}\,,\qquad s=1,2,3
\end{split}
\end{equation}
where:\footnote{
From now on we adopt the Einstein summation convention.
}
\begin{equation}\label{T0Ts}
\begin{aligned}
T_0(w) =& \frac{\vec{p}\cdot\vec{\nabla}}{2\gamma}w_0 +
\frac{\epsilon}{2}\vec{\nabla}\cdot\vec{w} + \Theta_\epsilon[V]w_0\,,\\
T_s(w) =& \frac{\vec{p}\cdot\vec{\nabla}}{2\gamma}w_s +
\frac{\epsilon}{2}\partial_s w_0 + \Theta_\epsilon[V]w_s +
\eta_{sjk}w_j p_k\,,\qquad s = 1,2,3\,,
\end{aligned}
\end{equation}
\begin{align*}
(\Theta_\epsilon(V)w)(r,p) =&
\frac{i}{\epsilon}(2\pi)^{-2}\int_{\mathbb{R}^2\times\mathbb{R}^2} \delta\tilde{V}(r,\xi)w(r,p')
e^{-i(p-p')\cdot\xi} d\xi dp'\,,\\
\delta \tilde{V}(r,\xi) =& V\bigg(r + \frac{\epsilon}{2}\xi\bigg) -
V\bigg(r - \frac{\epsilon}{2}\xi\bigg)\,,
\end{align*}
and:
\begin{equation}\label{gamma}
\gamma \equiv \frac{c}{\epsilon} = O(1)\quad(\epsilon\to 0)
\end{equation}
for the hypothesis \eqref{LSFS},
$\partial_s = \frac{\partial}{\partial_{r_s}}$ for $s = 1,2,3$, and $\eta_{sjk}$ denotes the only antisymmetric $3\times 3$ tensor which
is invariant for cyclic permutations of indexes and such that
$\eta_{123} = 1$ (in other words,  $\eta_{sjk} a_j b_k = (\vec{a}
\wedge \vec{b})_s$  for arbitrary $\vec{a},\,\vec{b}\in\mathbb{R}^3$).

\subsection{Equilibrium Wigner distribution in the diffusive case}
We are going to build two diffusive models for graphene by taking
moments of Eqs.\  \eqref{WDIFF2} and closing the resulting fluid-dynamic
equations by choosing the Wigner distribution $w$ equal to the
equilibrium distribution $g$, defined as a minimizer of a suitable
quantum entropy functional under the constraint of given moments. The
moments we choose are the following two:
\begin{equation}\label{npm}
n_\pm (r) = \int \left(w_0(r,p) \pm \frac{\vec{p}}{|\vec{p}|}\cdot\vec{w}(r,p)\right)\,dp\,.
\end{equation}
The $n_\pm$ are the so-called \emph{band densities}, that is, the
partial trace (w.r.t. $p$) of the quantum operators \emph{band
  projections} $\Pi_\pm$:\footnote{
Given a classical, complex hermitian matrix-valued symbol $a(r,p)$,
the expectation value of the quantum observable $A = \textnormal{Op}_\hbar(a)$ when the
system is in the state $S = \textnormal{Op}_\hbar(w)$ is:
$$\textnormal{Tr}(SA) = (2\pi\hbar)^{-2}\int_{\mathbb{R}^2\times\mathbb{R}^2}\textnormal{tr}(wa)\,drdp\,,$$
if $\textnormal{Tr}(SA)$ is finite.
}
\begin{align*}
\Pi_\pm =& \textnormal{Op}(P_\pm)\,,\quad
P_\pm(p) = \frac{1}{2}\left(\sigma_0 \pm \frac{p_s}{|\vec{p}|}\sigma_s\right)\,,\\
n_\pm(r) =& \textnormal{Tr}(\Pi_\pm S|r) = \int \textnormal{tr}(P_\pm(p)w(r,p))\,dp\,;
\end{align*}
here $S = \textnormal{Op}_\hbar(w)$ is the density operator which represents the state
of the system, and the matrices $P_\pm(p)$ are the projection
operators into the eigenspaces of the classical symbol $h$ of the
quantum Hamiltonian $H$, that is:
\begin{equation*}
H = \textnormal{Op}_\hbar(h)\,,\quad h(p) = E_+(p)P_+(p) + E_-(p)P_-(p)\,,
\end{equation*}
and $E_\pm(p)$ are the eigenvalues of $h$, that is, the energy bands
related to the Hamiltonian $H$:
\begin{equation*}
E_\pm(p) = \frac{|p|^2}{2m}\pm v_F|p|\,.
\end{equation*}
We define now the \emph{quantum entropy functional} with two equivalent formulations.
\begin{itemize}
\item Wigner function formulation:
\begin{equation}\label{Aw}
\mathcal{A}(w) = \int\textnormal{tr}(w(\mathcal{L}\textnormal{og}_\hbar(w) - 1 + h/k_B T))\,dr dp
\end{equation}
\item Density operators formulation:
\begin{equation}\label{AS}
\mathscr{A}(S) = \textnormal{Tr}(S(\log(S) - 1 + H/k_B T))\,
\end{equation}
\end{itemize}
where $\mathcal{L}\textnormal{og}_\hbar(w) = \textnormal{Op}_\hbar^{-1}\log\textnormal{Op}_\hbar(w)$ is the so-called
quantum logarithm\footnote{see \cite{Folland89, ZachosEtAl05} for
details.} of $w$, $k_B$ is the Boltzmann
constant, $T>0$ is the system temperature (which we assume constant)
and $H = \textnormal{Op}_\hbar(h)$ is the Hamiltonian \eqref{H}.

We define now the equilibrium Wigner distribution through the minimum
entropy principle. Let $\mathscr{W}$ the set of all Wigner functions
$w$ defined in $\mathbb{R}^2\times\mathbb{R}^2$ with complex hermitian matrix values;
for all $w\in\mathscr{W}$ we set:
\begin{equation}\label{wpm}
w_{\pm}(r,p) = w_0 (r,p) \pm
  \frac{\vec{p}}{|\vec{p}|}\cdot\vec{w} (r,p)\,.
\end{equation}
Moreover, let $\langle f \rangle = \int f\,dp$ for all $f(p)\in L^1(\mathbb{R}^2)$, and
let $n_+$, $n_-$ fluid-dynamic moments. We define the
\emph{Wigner distribution at local thermal equilibrium} related to
moments $n_+$, $n_-$ as the solution $g[n_+,n_-]$ of the problem:
\begin{equation}\label{CEM-D}
\mathcal{A}(g[n_+,n_-]) = \min_{\mathscr{W}}
\big\{\mathcal{A}(w)\,:\,\langle w_\pm \rangle = n_\pm\big\}\,.
\end{equation}
We are going to solve the problem \eqref{CEM-D} through the
Lagrange multipliers technique. Let us define the Lagrangian
functional in the following equivalent ways:
\begin{itemize}
\item with the Wigner functions formalism:
$$
\mathcal{L}(w,\xi_+,\xi_-) = \mathcal{A}(w) -
\int [\xi_+(n_+ - \langle w_+ \rangle) + \xi_-(n_- - \langle w_- \rangle)]\,dr
$$
for $w\in\mathscr{W}$ and $\xi_\pm(r)$ real functions;
\item with the density operators formalism:
\begin{equation*}\begin{split}
\mathscr{L}(S,\xi_+,\xi_-) =& \mathscr{A}(S) +
\textnormal{Tr}[\textnormal{Op}_\hbar(\xi_+ P_+ + \xi_- P_-)S]\\
&- (2\pi\hbar)^{-2}\int (\xi_+ n_+ + \xi_- n_-)\,dr
\end{split}\end{equation*}
for $S$ density operator and $\xi_\pm(r)$ real functions.
\end{itemize}
The Lagrange multipliers theory tells us (see \cite{DegondRinghofer02} for
details) that a necessary
condition for the $g[n_+,n_-]$ to solve the problem \eqref{CEM-D} is
 that the G\^ateaux derivative of $\mathscr{L}$
with respect to $S$ must vanish at
$S = \textnormal{Op}_\hbar(g[n_+,n_-])$, $\xi_+ = \xi_+^*$, $\xi_- = \xi_-^*$,
for a suitable choice of $\xi_+^*(r)$, $\xi_-^*(r)$ (Lagrange
multipliers):
\begin{equation}\label{eqLagr-diff}
\frac{\delta\mathscr{L}}{\delta S}(\textnormal{Op}_\hbar(g[n_+,n_-]),\xi_+^*,\xi_-^*) =
0\,.
\end{equation}
It can be proved (see \cite{DegondRinghofer03}) that the solution of
\eqref{eqLagr-diff} has the following form:
\begin{equation}\label{gndiff1}
\begin{split}
g[n_+,n_-] =& \mathcal{E}\textnormal{xp}_\hbar(-h_\xi)\,,\\
h_\xi =& \bigg(\frac{|p|^2}{2} + A\bigg)\sigma_0 +
(c|p| + B)\frac{\vec{p}}{|p|}\cdot\vec{\sigma}\,,
\end{split}
\end{equation}
where $\mathcal{E}\textnormal{xp}_\hbar(-h_\xi) \equiv \textnormal{Op}_\hbar^{-1}(\exp(\textnormal{Op}_\hbar(-h_\xi)))$ is the so-called quantum
exponential\footnote{see \cite{ Folland89, ZachosEtAl05} for details.}
of $-h_\xi$, while
$A=A(r) = (\xi_+^*(r) + \xi_-^*(r))/2$, $B=B(r) = (\xi_+^*(r) - \xi_-^*(r))/2$
 have to be determined in such a way that (recall \eqref{wpm}):
\begin{equation}\label{constraintQDE}
\langle g_\pm[n_+,n_-] \rangle (r) = n_\pm(r)\,,\qquad r\in\mathbb{R}^2\,.
\end{equation}

\subsection{Formal closure of the fluid equations in the diffusive case}
Let $n^\tau_+$, $n^\tau_-$ the moments of a solution $w = w^\tau$ of
\eqref{WDIFF2} with $g$ given by \eqref{gndiff1}, \eqref{constraintQDE}, and let:
\begin{equation}\label{T}
Tw = \sigma_0 T_0(w) + \sigma_s T_s(w)\,,
\end{equation}
for $w\in\mathscr{W}$ smooth enough function. We claim that:
\begin{equation}\label{hpQDE}
\langle (Tg[n_+^\tau,n_-^\tau])_\pm\rangle = 0\qquad\forall\tau>0\,.
\end{equation}
Indeed, it is immediate to verify
that Eq.~\eqref{hpQDE} is satisfied if
$g_0[n_+^\tau,n_-^\tau]$ is an even function of $p$ and
$\vec{g}[n_+^\tau,n_-^\tau]$ is an odd function of $p$;
as a matter of fact, $g[n_+^\tau,n_-^\tau]$
has this property, because of \eqref{gndiff1}. The proof of this claim
is quite similar to the proof of proposition $5.1$ at page 293 in
\cite{JSP10}: one only has to consider the operator $T$ given by
\eqref{T0Ts}, \eqref{T} instead of that one used in the paper and
consider $\mathcal{C}$ as the set of all the $p-$dependent
$2\times 2$ matrices with the parity structure:
$$(\textrm{even},\textrm{odd},\textrm{odd},\textrm{odd})$$
instead of:
$$(\textrm{even},\textrm{even},\textrm{odd},\textrm{even})\,.$$
The following (formal) result holds:
\begin{theorem}\label{teo} Let us suppose that:
$$n_\pm^\tau \to n_\pm \qquad\textrm{as }\tau\to 0\,;$$
then $n_+$, $n_-$ satisfy:
\begin{equation}\label{QDE-F}
\partial_t n_\pm = \langle (TTg[n_+,n_-])_\pm \rangle\,.
\end{equation}
\end{theorem}
We refer to \cite{JSP10} for the proof (the operator $T$
is slightly different there,
but the proof is still valid in our case).

The system \eqref{QDE-F} is closed because we already defined
(at least formally) the equilibrium distribution
$g[n_+^\tau,n_-^\tau]$. Nevertheless
it is very implicit, as the quantum exponential which appears in Eqs.\
\eqref{QDE-F} through Eq.\  \eqref{gndiff1} is very difficult to handle
with analytically and numerically. As anticipated, in the following
we will search for
an approximated but more explicit version of Eqs.\  \eqref{QDE-F}.

\subsection{First diffusive model: Explicit construction.}
In order to write \eqref{QDE-F} in an explicit way, we will exploit the
approximations we have done, that is, the semiclassical and the LSFS
hypothesis (given by \eqref{LSFS}). We will expand the equilibrium
distribution $g[n_+, n_-]$ at the first order in $\epsilon$,
neglecting $O(\epsilon^2)$ terms; to do this we start by
approximating the quantum exponential of an arbitrary classical symbol
with linear $\epsilon$-dependence. Let:
$$\{f,g\} =  \vec{\nabla}_r f \cdot\vec{\nabla}_p g - \vec{\nabla}_p f \cdot\vec{\nabla}_r g\,,$$
for all $f(r,p)$, $g(r,p)$ scalar smooth functions.
We apply here the general strategy for computing the semiclassical
expansion of the quantum exponential (see \cite{Jungel09, Jungel10}
for details).
Let $a = a_0\sigma_0 + \vec{a}\cdot\vec{\sigma}$,
$b = b_0\sigma_0 + \vec{b}\cdot\vec{\sigma}$ be arbitrary matrix
hermitian-valued classical symbols, and let us consider the function:
\begin{equation}\label{gbeta}
g_\epsilon(\beta) = \mathcal{E}\textnormal{xp}_\epsilon(\beta (a + \epsilon b))\,,\quad\beta\in\mathbb{R}\,.
\end{equation}
Let us recall the definition of the so-called \emph{Moyal product}:
\begin{equation}\label{moyal}
f_1\#_\epsilon f_2 = \textnormal{Op}_\epsilon^{-1}(\textnormal{Op}_\epsilon(f_1)\textnormal{Op}_\epsilon(f_2))
\end{equation}
between arbitrary classical symbols $f_1$, $f_2$. It is known
\cite{DegondMehatsRingh05} that the Moyal product has a semiclassical expansion:
\begin{equation*}
\#_\epsilon = \sum_{n=0}^\infty \epsilon^n\#^{(n)}\,,
\end{equation*}
and the first three terms of this expansion (the only terms needed in
this work) are:
\begin{equation}\label{moy012}
\begin{aligned}
f_1\#^{(0)} f_2 &= f_1 f_2\,,\\
f_1\#^{(1)} f_2 &= \frac{i}{2}\left(\partial_{r_s}f_1 \partial_{p_s}f_2
  - \partial_{p_s}f_1 \partial_{r_s}f_2\right)\,,\\
f_1\#^{(2)} f_2 &= -\frac{1}{8}\left(\partial^2_{r_j
    r_s}f_1 \partial^2_{p_j p_s}f_2
- 2 \partial^2_{r_j p_s}f_1 \partial^2_{p_j r_s}f_2
+ \partial^2_{p_j p_s}f_1 \partial^2_{r_j r_s}f_2\right)\,.
\end{aligned}
\end{equation}
Now let us differentiate with respect to $\beta$ the function
$g_\epsilon(\beta)$ given by \eqref{gbeta}. By using the definition
\eqref{moyal} of the Moyal product we obtain:
\begin{equation}\label{dgbeta}
\partial_\beta g_\epsilon(\beta) = \frac{1}{2}((a+\epsilon b)\#_\epsilon g_\epsilon(\beta) +
g_\epsilon(\beta)\#_\epsilon (a+\epsilon b))\,,
\end{equation}
and $g_\epsilon(0) = \sigma_0$. So by expanding the expressions in
\eqref{dgbeta} in powers of $\epsilon$ we find:
\begin{equation}\label{dgbeta0}
\partial_\beta g^{(0)}(\beta) = \frac{1}{2}(g^{(0)}(\beta)a + ag^{(0)}(\beta))\,,
\end{equation}

\begin{equation}\label{dgbeta1}
\begin{aligned}
\partial_\beta g^{(1)}(\beta) =& \frac{1}{2}(g^{(1)}(\beta)a +
ag^{(1)}(\beta)) +
\frac{1}{2}(g^{(0)}(\beta)b + bg^{(0)}(\beta))\\
&+\frac{1}{2}(g^{(0)}(\beta)\#^{(1)}a + a \#^{(1)} g^{(0)}(\beta))\,,
\end{aligned}
\end{equation}
with the initial conditions:
\begin{equation}\label{dgbeta-init} g^{(0)}(0) = \sigma_0\,,\qquad g^{(1)}(0) = 0\,. \end{equation}
The equations \eqref{dgbeta0}, \eqref{dgbeta1} with
the initial conditions \eqref{dgbeta-init} can be exactly
solved in this order to obtain the $O(\epsilon^2)-$approximation of
$\mathcal{E}\textnormal{xp}_\epsilon(a) = g_\epsilon(1)$: in fact, each equation is a linear
ODE with constant coefficients. It is easy to find the leading term in the
expansion of $g_\epsilon(\beta)$:
\begin{equation}\label{gbeta0}
g^{(0)}(\beta) = \exp(\beta a) = e^{\beta a_0}\left(\cosh(\beta|\vec{a}|)\sigma_0 +  \sinh(\beta|\vec{a}|)\frac{\vec{a}}{|\vec{a}|}\cdot\vec{\sigma}\right)\,.
\end{equation}
We now have to explicitly compute the first order correction of
$g_\epsilon(\beta)$ from \eqref{dgbeta1}; to this aim, it is
useful to employ some properties of the Pauli matrices. It is easy to
verify that, for $a$, $b$ arbitrary hermitian matrix-valued classical
symbols holds:
\begin{equation}\label{pauli-moy}
\begin{aligned}
\frac{1}{2}(a\#^{(k)}b + b\#^{(k)}a) =& (a_0\#^{(k)}b_0 +
\vec{a}\cdot^{\#^{(k)}}\vec{b})\sigma_0 \\
& + (a_0\#^{(k)}\vec{b} +
b_0\#^{(k)}\vec{a})\cdot\vec{\sigma}\quad\textrm{for even $k$,}\\
\frac{1}{2}(a\#^{(k)}b + b\#^{(k)}a) =& i(\vec{a}\wedge^{\#^{(k)}}\vec{b})\cdot\vec{\sigma}\quad\textrm{for odd $k$,}
\end{aligned}
\end{equation}
where we defined:
$$\vec{a}\cdot^{\#^{(k)}}\vec{b} = a_s \#^{(k)}b_s\,,\quad
(\vec{a}\wedge^{\#^{(k)}}\vec{b})_j = \eta_{jst}a_s \#^{(k)} b_t\,.$$
The relations \eqref{pauli-moy} allow us to reduce the calculus of the matrix
 $g^{(1)}(\beta)$ to that of its Pauli components; if fact, due to
 \eqref{pauli-moy}, \eqref{dgbeta1} becomes:
\begin{equation}\label{dgbeta1pauli}
\begin{aligned}
\partial_\beta g^{(1)}_0(\beta) =& a_0 g^{(1)}_0(\beta) +
\vec{a}\cdot\vec{g}^{(1)}(\beta) + b_0 g^{(0)}_0(\beta) +
\vec{b}\cdot\vec{g}^{(0)}(\beta)\\
\partial_\beta \vec{g}^{(1)}(\beta) =&  a_0 \vec{g}^{(1)}(\beta) +
\vec{a}g^{(1)}_0(\beta) + b_0 \vec{g}^{(0)}(\beta) +
\vec{b}g^{(0)}_0(\beta) + i\vec{a}\wedge^{\#^{(1)}}\vec{g}^{(0)}(\beta)
\end{aligned}
\end{equation}
In order to solve \eqref{dgbeta1pauli}, let us consider the
homogeneous problem:
\begin{equation}\label{hompb}
\begin{aligned}
\partial_\beta x_0(\beta) =& a_0 x_0(\beta) +
\vec{a}\cdot\vec{x}(\beta)\qquad\beta>0\,,\\
\partial_\beta \vec{x}(\beta) =&  a_0 \vec{x}(\beta) +
\vec{a}x_0(\beta) \qquad\beta>0\,.
\end{aligned}
\end{equation}
The problem \eqref{hompb} can be solved with elementary techniques, finding
that the vector $X(\beta) = [x_0(\beta), \vec{x}(\beta)]$ is given by:
$$X(\beta) = S_a(\beta)X(0) \qquad\beta>0\,,$$
with the semigroup operator $S_a(\beta)$ defined by:
\begin{equation}\label{Sabeta}
S_a(\beta) = e^{\beta a_0}\begin{pmatrix}
\cosh(\beta|\vec{a}|) &
\sinh(\beta|\vec{a}|)\vec{\alpha}^T\\
\sinh(\beta|\vec{a}|)\vec{\alpha} &
(\cosh(\beta|\vec{a}|)-1)\vec{\alpha}\otimes\vec{\alpha}+I_{3\times 3}
\end{pmatrix}
\end{equation}
with $\vec{\alpha} \equiv \vec{a}/|\vec{a}|$, and $\vec{\alpha}^T$
denotes the transpose of the vector $\vec{\alpha}$. Now the
semigroup theory allows us to write the solution of
\eqref{dgbeta1pauli}:
\begin{equation}\label{gbeta10-f}
\begin{aligned}
g^1_0(\beta) =& \int_0^\beta e^{(\beta-\lambda)a_0}
\cosh((\beta - \lambda)|\vec{a}|)Y_0(\lambda)\,d\lambda \\
& +\int_0^\beta e^{(\beta-\lambda)a_0}\sinh((\beta-\lambda)|\vec{a}|)\frac{\vec{a}}{|\vec{a}|}\cdot \vec{Y}(\lambda)\,d\lambda\,,
\end{aligned}
\end{equation}
\begin{equation}\label{gbeta1v-f}
\begin{aligned}
\vec{g}^1(\beta)
=& \int_0^\beta e^{(\beta-\lambda)a_0}
\sinh((\beta-\lambda)|\vec{a}|)\frac{\vec{a}}{|\vec{a}|}Y_0(\lambda)
\,d\lambda\\
& +\int_0^\beta e^{(\beta-\lambda)a_0}
\left([\cosh((\beta-\lambda)|\vec{a}|)-1]\frac{\vec{a}\otimes\vec{a}}{|\vec{a}|^2}+I_{3\times
    3}\right)\vec{Y}(\lambda) \,d\lambda\,,
\end{aligned}
\end{equation}
\begin{equation}\label{gbeta1++}
\begin{split}
Y_0(\lambda) =& b_0 g^{(0)}_0(\lambda) +
\vec{b}\cdot\vec{g}^{(1)}(\lambda)\,,\\
\vec{Y}(\lambda) =& b_0 \vec{g}^{(0)}(\lambda) +
\vec{b}g^{(1)}_0(\lambda)+
i\vec{a}\wedge^{\#^{(1)}}\vec{g}^{(0)}(\lambda)\,;
\end{split}
\end{equation}

finally, from \eqref{gbeta0}, \eqref{gbeta10-f}, \eqref{gbeta1v-f},
\eqref{gbeta1++} we obtain:
\begin{equation}\label{gbeta10}
\begin{aligned}
&g^{(1)}_0(\beta)\\
 =& \beta e^{\beta a_0}\left(\cosh(\beta|\vec{a}|)b_0+
\sinh{(\beta|\vec{a}|)}\frac{\vec{a}\cdot\vec{b}}{|\vec{a}|}\right)\\
&+\beta e^{\beta a_0}\frac{\sinh(\beta|\vec{a}|)-\beta|\vec{a}|\cosh(\beta|\vec{a}|)}{4 \beta|\vec{a}|^3}\eta_{jks}\{a_j,a_k\}a_s\,,
\end{aligned}
\end{equation}
\begin{equation}\label{gbeta1v}
\begin{aligned}
&\vec{g}^{(1)}(\beta) \\
=& \beta e^{\beta a_0}\bigg(
\sinh(\beta|\vec{a}|)b_0+
\cosh{(\beta|\vec{a}|)}\frac{\vec{a}\cdot\vec{b}}{|\vec{a}|}
-\frac{\sinh(\beta|\vec{a}|)}{4|\vec{a}|^2}\eta_{jks}\{a_j,a_k\}a_s
\bigg)\frac{\vec{a}}{|\vec{a}|}\\
&+\beta e^{\beta a_0}\frac{\sinh(\beta|\vec{a}|)}{\beta|\vec{a}|}
\left(\frac{\vec{a}\wedge\vec{b}}{|\vec{a}|}\right)
\wedge\frac{\vec{a}}{|\vec{a}|}+\\
&+\beta e^{\beta a_0}\frac{\beta|\vec{a}|\sinh(\beta|\vec{a}|) - \cosh(\beta|\vec{a}|) +
  1}{2 \beta|\vec{a}|^3}a_j\{a_j,\vec{a}\}\wedge\frac{\vec{a}}{|\vec{a}|}\\
&+\beta e^{\beta a_0}\frac{\beta|\vec{a}|\cosh(\beta|\vec{a}|) - \sinh(\beta|\vec{a}|)}{2 \beta|\vec{a}|^2}\{a_0,\vec{a}\}\wedge\frac{\vec{a}}{|\vec{a}|}
\,.
\end{aligned}
\end{equation}

So we have explicitly computed the first-order semiclassical expansion
of $g_\epsilon(\beta) = \mathcal{E}\textnormal{xp}_\epsilon(\beta(a + \epsilon b))$.
We point out that in the scalar case the odd order terms in the
semiclassical expansion of the quantum exponential are zero, while
this does not happen in the spinorial case, due to the
noncommutativity of the matrix product, which increases much the
complexity in computation with respect to the scalar case.

The next step is the computation of $g[n_+,n_-]$ given by
\eqref{gndiff1}, \eqref{constraintQDE}. To achieve this goal,
we substitute in \eqref{gbeta0}, \eqref{gbeta10}, \eqref{gbeta1v}
$\beta = 1$ and $a + \epsilon b = -h_\xi$; this means, due to
\eqref{gndiff1}, \eqref{gamma}:
$$-a = \bigg(\frac{|p|^2}{2} + A\bigg)\sigma_0 +
B\frac{\vec{p}}{|p|}\cdot\vec{\sigma}\,,\quad
-b = \gamma\vec{p}\cdot\vec{\sigma}\,.$$
By making all the straightforward computations needed, we finally
find:
\begin{equation}\label{gn}
\begin{aligned}
g_0[n_+,n_-] =& \frac{e^{-|\vec{p}|^2/2}}{2\pi}\left\{
n_0 + \epsilon\gamma\left(\sqrt\frac{\pi}{2} -
  |\vec{p}|\right)n_\sigma\right\} + O(\epsilon^2)\,,\\
\vec{g}[n_+,n_-] =&\frac{e^{-|\vec{p}|^2/2}}{2\pi}\bigg\{
n_\sigma\frac{\vec{p}}{|\vec{p}|} +
\epsilon\gamma\left(\sqrt\frac{\pi}{2} -
    |\vec{p}|\right)n_0\frac{\vec{p}}{|\vec{p}|}
+ \epsilon\vec{F}\wedge \frac{\vec{p}}{|\vec{p}|^2}\bigg\} + O(\epsilon^2)\,,
\end{aligned}
\end{equation}
with:
\begin{equation}\label{F}
\vec{F} = -\frac{1}{2}\left[\vec{\nabla}|n_\sigma| -
\log\sqrt\frac{n_0 + |n_\sigma|}{n_0 - |n_\sigma|}\,\vec{\nabla}\left(
n_0 + \sqrt{n_0^2 - n_\sigma^2}\right)\right]\,,
\end{equation}
and:
\begin{equation}\label{n0nsigma}
\begin{aligned}
n_0 =& \frac{1}{2}(n_+ + n_-)\qquad\textrm{particle density,}\\
n_\sigma =& \frac{1}{2}(n_+ - n_-)\qquad\textrm{pseudo-spin polarization.}\\
\end{aligned}
\end{equation}

At this point, using \eqref{gn}, \eqref{F}, \eqref{n0nsigma} to
explicitly compute the terms in \eqref{QDE-F} up to $O(\epsilon^2)$,
we obtain the desired QDE model:
\begin{equation}\label{QDE}
\begin{aligned}
\partial_t n_0 =& \frac{\Delta}{4\gamma^2}\left[
n_0 + \frac{\epsilon\hat{\gamma}}{2}n_\sigma\right] +
\frac{\vec{\nabla}}{2\gamma}\cdot(n_0\vec{\nabla}V) +
O(\epsilon^2)\,,\\
\partial_t n_\sigma =& \frac{\Delta}{4\gamma^2}\left[
n_\sigma + \frac{\epsilon\hat{\gamma}}{2}n_0\right]\\
&-\frac{\vec{\nabla}V}{2\gamma}\cdot\left[
\vec{\nabla}n_\sigma + \epsilon\hat{\gamma}\vec{\nabla}n_0 +
\frac{1}{2}\left(1 + \frac{n_0}{\sqrt{n_0^2 - n_\sigma^2}}\right)\frac{
\vec{\nabla}n_0\wedge\vec{\nabla}n_\sigma}{\sqrt{n_0^2 - n_\sigma^2}}
\right]\\
&-\frac{|\vec{\nabla}V|^2}{2}[n_\sigma+\epsilon\hat{\gamma}(\Gamma -1) n_0]
-\frac{3}{4\gamma}\Delta V[n_\sigma+\epsilon\hat{\gamma} n_0]+
O(\epsilon^2)\,,
\end{aligned}
\end{equation}
with:
\begin{equation}\label{Gamma}
\Gamma = \frac{1}{2\pi}\int_0^{\infty}
e^{-\rho^2/2}\rho\log\rho\,d\rho\,,\qquad \hat{\gamma} = \gamma\sqrt\frac{\pi}{2}
\,.
\end{equation}

\section{A second diffusive model for graphene}
We are going to build another diffusive model for quantum transport in
graphene, starting again from the Wigner equations in diffusive
scaling \eqref{WDIFF2}, considering the same fluid-dynamic moments
$n_\pm$ of the Wigner distribution $w(r,p,t)$ and taking again as the
equilibrium distribution the one given in \eqref{gndiff1}, \eqref{constraintQDE};
however, we will make
stronger assumptions than \eqref{LSFS}, which will allow us to
consider also $O(\epsilon^2)-$terms in the fluid equations.
\subsection{Assumptions}
We make the semiclassical hypothesis $\epsilon\ll 1$ like in the
previous model, and another hypothesis, stronger than
\eqref{LSFS}, which we call \emph{Strongly Mixed State} (SMS):
\begin{equation}\label{SMS}
c = O(\epsilon)\,,\qquad B = O(\epsilon)\,.
\end{equation}
(Recall the definitions \eqref{DSCAL-3}, \eqref{gndiff1} of $c$ and
$B$, respectively). These further assumptions are necessary to overcome
the computational difficulties arising from the spinorial nature of
the problem: without these hypothesis, it would be hard to compute
the second order expansion of the equilibrium distribution.

We will see that the two approximations will result in the fact:
\begin{equation}\label{SMS2}
\left|\frac{n_+ - n_-}{n_+ + n_-}\right| = \left|\frac{n_\sigma}{n_0}\right| = O(\epsilon)\,.
\end{equation}
This means, from a physical point of view, that the charge carriers
have approximately the same probability of being found in the
conduction band or in the valence band of the energy spectrum,
or equivalently, there is little
difference (with respect to the total charge density) between the
electron density and the hole density.
Analytically, the main consequence of the SMS hypothesis \eqref{SMS}
will be the decoupling of the modified
Hamiltonian in a scalar part of order 1 and a spinorial perturbation
of order $\epsilon$; this fact will be very useful in computations.

\subsection{Second diffusive model: Explicit construction.}
For the sake of simplicity let us redefine $B\mapsto \epsilon B$ and consider $B = O(1)$.\\
Under our hypothesis, the classical symbol of the modified Hamiltonian becomes:
$$h_\xi = \left(\frac{|p|^2}{2} + A\right)\sigma_0 + \epsilon(\gamma|p|+B)\frac{\vec{p}}{|p|}\cdot\vec{\sigma}\,;
$$
that is, as anticipated, the modified Hamiltonian $h_\xi$ decouples in
a \emph{scalar} part of order $O(1)$ and a \emph{spinorial} part of
order $O(\epsilon)$, so that $h_\xi$ can be seen as a small perturbation of
a scalar Hamiltonian. We are going to see that this fact leads to
remarkable simplifications in computations.

In order to explicitly compute the equilibrium distribution,
let us recall \eqref{gbeta0}, \eqref{dgbeta1pauli}. In this case we have:
\begin{equation}\label{abQDE2}
-a = \left(\frac{|p|^2}{2} + A\right)\sigma_0\,,\qquad
-b = (\gamma|p|+B)\frac{\vec{p}}{|p|}\cdot\vec{\sigma}\,,
\end{equation}
so \eqref{gbeta0} becomes the \emph{scalar} function:
\begin{equation}\label{gbeta0-bis}
g^{(0)}(\beta) = e^{\beta a}\sigma_0\,,
\end{equation}
and \eqref{dgbeta1pauli} takes the form:
\begin{equation*}\label{dgbeta-bis}
\partial_\beta g^{(1)}(\beta) = a g^{(1)}(\beta) +
g^{(0)}(\beta)b\qquad\beta > 0\,;
\end{equation*}
then the solution is:
\begin{equation}\label{gbeta1-f-bis}
g^{(1)}(\beta) = -\beta e^{-\beta(|p|^2/2 + A)}(B + \gamma |p|)\frac{\vec{p}}{|p|}\cdot\vec{\sigma}\,.
\end{equation}
Now we can compute also the second-order correction to
$g_\epsilon(\beta)$, thanks to the approximations we have done.
From \eqref{dgbeta} we obtain:
\begin{equation*}
\partial_\beta g^{(2)}(\beta) = a g^{(2)}(\beta) +
a\#^{(2)}g^{(0)}(\beta) + \frac{1}{2}(b g^{(1)}(\beta) + g^{(1)}(\beta)b)
\end{equation*}
so one finds that:\footnote{
Here $\mathcal{E}\textnormal{xp}(\beta a)^{(2)}$ is the second order term in the
semiclassical expansion of $\mathcal{E}\textnormal{xp}_\epsilon(\beta a)$.
}
\begin{equation}\label{gbeta2-f}
g^{(2)}(\beta) = \mathcal{E}\textnormal{xp}(\beta a)^{(2)} + \frac{\beta^2
  |\vec{b}|^2}{2}e^{\beta a}\,.
\end{equation}
So from \eqref{abQDE2}, \eqref{gbeta0-bis}, \eqref{gbeta1-f-bis}, \eqref{gbeta2-f} we
find:
\begin{equation}\label{gbeta-bis}
\begin{aligned}
g_0(\beta) =& \tilde{g}(\beta) +
\frac{\epsilon^2}{2}\beta^2 e^{-\beta(|p|^2/2 + A)} (B + \gamma
|p|)^2 + O(\epsilon^3)\,,\\
\vec{g}(\beta) =& -\epsilon\beta e^{-\beta(|p|^2/2 + A)}(B + \gamma
|p|)\frac{\vec{p}}{|p|} + O(\epsilon^3)\,,
\end{aligned}
\end{equation}
where $\tilde{g}(\beta)$ is the second-order approximation of the
quantum exponential of the scalar symbol $\beta a$:
\begin{equation}\label{gtildeQDE}
\mathcal{E}\textnormal{xp}_\epsilon(\beta a) = \tilde{g}(\beta) + O(\epsilon^3)\,.
\end{equation}
It is convenient to work from now on with the moments $n_0$,
$n_\sigma$ given by \eqref{n0nsigma}
instead of $n_\pm$; for this reason we will write $g[n_0,n_\sigma]$
instead of $g[n_+,n_-]$, being no possibility of misunderstanding in
this. The constraint \eqref{constraintQDE} can so be rewritten in terms
of $g[n_0,n_\sigma]$ in the following way\footnote{
We adopt here the obvious notation: $w_\sigma = \vec{p}\cdot \vec{w}/|\vec{p}|$ for all
$w$ spinorial Wigner functions.}:
\begin{equation}\label{constraintQDE2}
\langle g_0[n_0,n_\sigma] \rangle = n_0\,,\qquad\langle g_\sigma[n_0,n_\sigma]\rangle = n_\sigma\,;
\end{equation}
By imposing the constraint on the moments and
 making all the straightforward computations needed, we obtain the
following result:
\begin{equation}\label{gQDE2}
\begin{split}
g_0[n_0,n_\sigma] =&\tilde{g}\left[n_0 -
  \frac{n_0}{2}\left(\epsilon^2\gamma^2(2 - \pi/2) + (n_\sigma/n_0)^2\right)
\right]\\
&+ \frac{n_0}{4\pi}e^{-|p|^2/2}\left[
\epsilon\gamma\left(\sqrt\frac{\pi}{2} - |p|\right) + \frac{n_\sigma}{n_0}
\right]^2 + O(\epsilon^3)\,,\\
\vec{g}[n_0,n_\sigma] =& \epsilon\frac{n_0}{2\pi}e^{-|p|^2/2}\left[\frac{n_\sigma}{\epsilon n_0} +
\gamma\left(\sqrt\frac{\pi}{2} - |p|\right)\right]\frac{\vec{p}}{|\vec{p}|} + O(\epsilon^3)\,,
\end{split}
\end{equation}
where $\tilde{g}[n]$ is the $O(\epsilon^3)-$approximation of the
scalar quantum Maxwellian with $\langle\tilde{g}[n]\rangle = n$, that is:
\begin{equation*}
\tilde{g}[n] = \frac{n}{2\pi}e^{-|p|^2/2}\left[1+\frac{\epsilon^2}{24}
\vec{\nabla}\cdot\left((I - \vec{p}\otimes\vec{p})\vec{\nabla}\log n\right)\right]
\end{equation*}
for an arbitrary positive scalar function $n(r)$ and $I$ is the $3\times 3$ identity matrix.

We point out that, during the computations, we find (see the Appendix):
\begin{equation}\label{SMS2-bis}
\epsilon n_0\left(B + \gamma\sqrt\frac{\pi}{2}\right) = -n_\sigma +
O(\epsilon^3)\,,
\end{equation}
so the constraint \eqref{SMS2} is satisfied.

Finally, by using Eqs.\  \eqref{gQDE2} to compute the terms in
\eqref{QDE-F} and
making all the long but straightforward computations needed,
we obtain the second semiclassical quantum diffusive model we were looking for:
\begin{equation}\label{QDE2}
\begin{split}
\partial_t n_0 =& D_{00}^\epsilon\Delta n_0 + D_{0\sigma}^\epsilon\Delta n_\sigma +
\frac{1}{2\gamma}\vec{\nabla}\cdot\left(n_0\vec{\nabla}(V + V_B)\right) + O(\epsilon^3)\,,\\
\partial_t n_\sigma =& D_{\sigma 0}^\epsilon\Delta n_0 + D_{\sigma \sigma}^\epsilon\Delta n_\sigma
-\frac{1}{2\gamma}\vec{\nabla}\cdot\left(\left(n_\sigma + \epsilon\hat{\gamma}n_0\right)\vec\nabla V\right)\\
& + \frac{1}{2}|\vec\nabla V|^2 \left[\Gamma n_\sigma + \epsilon\hat{\gamma}
\left(2 + \Gamma\right)n_0\right]
 +\frac{1}{4\gamma}\Delta V\left[n_\sigma + \epsilon\hat{\gamma}n_0\right] +
O(\epsilon^3)\,,
\end{split}
\end{equation}
where $\Gamma$, $\hat{\gamma}$ are given by \eqref{Gamma}, $V_B$ is the so-called
\emph{Bohm potential}:
\begin{equation}\label{Bohm}
V_B = -\frac{\epsilon^2}{6}\frac{\Delta\sqrt{n_0}}{\sqrt{n_0}}\,,
\end{equation}
and we defined the following constants:
\begin{align*}
D_{00}^\epsilon =& \frac{1}{4\gamma^2} + \frac{\epsilon^2}{4}(2 - \gamma(4-\pi))\,,\quad
D_{0\sigma}^\epsilon = D_{\sigma 0}^\epsilon = \frac{\epsilon}{8\gamma}\sqrt\frac{\pi}{2}\,,\quad
D_{\sigma\sigma}^\epsilon = \frac{1}{4\gamma^2}\,.
\end{align*}
We will dedicate the remaining part of this paper to the
construction of hydrodynamic models for graphene.

\section{A first hydrodynamic model for graphene}
In the following of this paper we will build two hydrodynamic models for quantum electron transport in graphene following a strategy similar to that one employed in the construction of the diffusive models \eqref{QDE}, \eqref{QDE2}.
\subsection{Wigner equations in hydrodynamic scaling}
Let us reconsider the Wigner equations \eqref{WE2}
 and perform the following hydrodynamic scaling:
\begin{equation}\label{HSCAL-1}
r\mapsto \hat{r} r\,,\quad t\mapsto \hat{t} t\,,\quad p\mapsto \hat{p}
p\,,\quad V\mapsto \hat{V} V\,,
\end{equation}
with $\hat{r}$, $\hat{t}$, $\hat{p}$, $\hat{V}$ satisfying:
\begin{equation}\label{HSCAL-2}
\frac{1}{\hat{t}} = \frac{2 v_F \hat{p}}{\hbar} = \frac{\hat{V}}{\hat{r} \hat{p}}\,,
\quad \hat{p} = \sqrt{m k_B T}\,;
\end{equation}
moreover, let us define the \emph{semiclassical parameter} $\epsilon$
and the \emph{hydrodynamic parameter} $\tau$ as:
\begin{equation}\label{HSCAL-3}
\epsilon = \frac{\hbar}{\hat{r} \hat{p}}\,,\quad \tau = \frac{\tau_c}{\hat{t}}\,.
\end{equation}
Let $c$ (scaled Fermi speed) be given again by \eqref{DSCAL-3}. We make
the same assumptions as in the first diffusive model, that is, the
semiclassical hypothesis $\epsilon \ll 1$ and the LSFS hypothesis
\eqref{LSFS}. Let $\gamma$ be defined as in \eqref{gamma}. If we perform the
scaling \eqref{HSCAL-1} -- \eqref{HSCAL-3} on \eqref{WE2} under
the assumptions we have made, we obtain the following scaled Wigner system:
\begin{equation}\label{WHYDRO2}
\begin{split}
\partial_t w_0 + \frac{\vec{p}\cdot\vec{\nabla}}{2\gamma}w_0 +\frac{\epsilon}{2}\vec{\nabla}\cdot\vec{w} + \Theta_\epsilon[V]w_0 =& \frac{g_0[n_0,\vec{n},\vec{J}] - w_0}{\tau}\,,\\
\partial_t \vec{w} + \frac{\vec{p}\cdot\vec{\nabla}}{2\gamma}\vec{w}+ \frac{\epsilon}{2}\vec{\nabla}w_0 + \vec{w}\wedge\vec{p} + \Theta_\epsilon[V]\vec{w} =& \frac{\vec{g}[n_0,\vec{n},\vec{J}] - \vec{w}}{\tau}\,.
\end{split}
\end{equation}
$g$ is the quantum thermal equilibrium Wigner distribution, which will
be different from the diffusive case because we are going to choose a
different set of moments.

\subsection{Equilibrium Wigner distribution in the hydrodynamic case}
The moments we choose are the following six:
\begin{equation*}
\begin{split}
n_s =& \int w_s\,dp\qquad s = 0,1,2,3\,,\\
J_k =& \int p_k w_0\,dp\qquad k = 1,2\,.
\end{split}
\end{equation*}
$n_0$ is the \emph{particle density}, $\vec{n} = (n_1, n_2, n_3)$ is the \emph{spin vector},
$\vec{J} = (J_1,J_2,0)$ is the \emph{flow vector}. Note that the flow vector has only two components because graphene is a two-dimensional object.

We choose \eqref{Aw}, \eqref{AS}
as quantum entropy like in the diffusive case,
and we define the equilibrium distribution $g[n_0,\vec{n},\vec{J}]$
again through the MEP. Let $n_0$, $\vec{n}$, $\vec{J}$ denote given moments.
We define the \emph{Wigner distribution at local thermal equilibrium}
related to moments $n_0$, $\vec{n}$, $\vec{J}$ the solution
$g[n_0,\vec{n},\vec{J}]$ of the problem:
\begin{equation}\label{CEM-H}
\mathcal{A}(g[n_0,\vec{n},\vec{J}]) = \min
\big\{\mathcal{A}(w)\,:\,w\in\mathscr{W}\,,\,\,\langle w_0\rangle = n_0\,,\,\,
\langle\vec{w}\rangle = \vec{n}\,,\,\,\langle\vec{p}w_0 \rangle = \vec{J}\big\}\,.
\end{equation}
In order to solve this problem, let us
introduce (similarly to the diffusive case)
the Lagrangian functional:
$$\begin{aligned}
\tilde{\mathscr{L}}(S,\xi_0^0,(\xi_s^0)_{s=1,2,3},(\xi_0^k)_{k=1,2}) =& \mathscr{A}(S) +
\textnormal{Tr}(\textnormal{Op}_\hbar[(\xi_0^0 + p_k \xi_0^k)\sigma_0 + \xi_s^0\sigma_s]S)\\
&- 2(2\pi\hbar)^{-2}\int \big[\xi_0^0n_0 + \xi_s^0n_s + \xi_0^k J_k\big]\,dr\,,
\end{aligned}$$
defined for $S$ density operator and
$\xi_0^0(r),(\xi_s^0(r))_{s=1,2,3},(\xi_0^k(r))_{k=1,2}$ real
functions, with $\mathscr{A}(S)$ given by \eqref{AS}. Again,
the Lagrange multipliers theory tells us (see \cite{DegondRinghofer02} for
details) that a necessary
condition for the $g[n_0,\vec{n},\vec{J}]$ to solve the problem \eqref{CEM-H} is
 that the G\^ateaux derivative of $\tilde{\mathscr{L}}$
with respect to $S$ must vanish when:
$$\begin{aligned}
S =& \hat{S} \equiv \textnormal{Op}_\hbar(g[n_0,\vec{n},\vec{J}])\,,\\
\xi_0^0(r) =& \hat{\xi}_0^0(r)\,,\\
\xi_s^0(r) =& \hat{\xi}_s^0(r)\qquad s = 1,2,3\,,\\
\xi_0^k(r) =& \hat{\xi}_0^k(r)\qquad k = 1,2\,,
\end{aligned}$$
for a suitable choice of
$\hat{\xi}_0^0(r),\,(\hat{\xi}_s^0(r))_{s=1,2,3},\,(\hat{\xi}_0^k(r))_{k=1,2}$
(Lagrange multipliers):
\begin{equation}\label{eqLagr-hydro}
\frac{\delta\tilde{\mathscr{L}}}{\delta S}(\hat{S},
\hat{\xi}_0^0,(\hat{\xi}_s^0)_{s=1,2,3},(\hat{\xi}_0^k)_{k=1,2}) =
0\,.
\end{equation}
It can be proved (see \cite{DegondRinghofer03}) that the
solution of \eqref{eqLagr-hydro} has the following form:
\begin{equation}\label{ghydro}
\begin{split}
g[n_0,\vec{n},\vec{J}] =& \mathcal{E}\textnormal{xp}(-h_\xi)\,,\\
h_\xi =& \bigg(\frac{|p|^2}{2} + p_k\xi_0^k + \xi_0^0\bigg)\sigma_0 +
(\xi_s^0 + c p_s)\sigma_s\,,
\end{split}
\end{equation}
with $\xi_0^0(r)$, $(\xi_s^0(r))_{s=1,2,3}$, $(\xi_0^k(r))_{k=1,2}$
Lagrange multipliers to be determined in such a way that:
\begin{equation}\label{constraintsQHE}
\langle g_0[n_0,\vec{n},\vec{J}]\rangle (r) = n_0 (r)\,,\,\,
\langle\vec{g}[n_0,\vec{n},\vec{J}]\rangle (r) = \vec{n}(r)\,,\,\,
\langle\vec{p}g_0[n_0,\vec{n},\vec{J}]\rangle (r) = \vec{J}(r)\,,
\end{equation}
for $r\in\mathbb{R}^2$.

\subsection{Formal closure of the fluid equations in the hydrodynamic case.}
The following theorem holds:
\begin{theorem}\label{teo2} Let $n_0^\tau$, $\vec{n}^\tau$, $\vec{J}^\tau$ the moments of a solution $w^\tau$ of Eqs.\  \eqref{WHYDRO2}. If $n_0^\tau\to n_0$, $\vec{n}^\tau\to\vec{n}$,
$\vec{J}^\tau\to\vec{J}$ as $\tau\to 0$, then the limit moments $n_0$, $\vec{n}$, $\vec{J}$
satisfy:
\begin{equation}\label{QHE-F}
\begin{split}
\partial_t n_0 + \frac{\vec{\nabla}}{2\gamma}\cdot\vec{J} + \frac{\epsilon}{2}\vec{\nabla}\cdot\vec{n} =& 0\\
\partial_t\vec{n} + \frac{\vec{\nabla}}{2\gamma}\cdot \langle\vec{p}\otimes\vec{g}\rangle +
\frac{\epsilon}{2}\vec{\nabla}n_0 + \langle\vec{g}\wedge\vec{p}\rangle =& 0\\
\partial_t\vec{J} + \frac{\vec{\nabla}}{2\gamma}\cdot\left(\frac{\vec{J}\otimes\vec{J}}{n_0} + \mathscr{P}\right) + \frac{\epsilon}{2}\vec{\nabla}\cdot \langle\vec{p}\otimes\vec{g}\rangle + n_0\vec{\nabla}V =& 0
\end{split}
\end{equation}
where:
$$\mathscr{P} = \langle (\vec{p} - \vec{J}/n_0)\otimes(\vec{p} - \vec{J}/n_0)g_0 \rangle$$
is the so-called \emph{quantum stress tensor}.
\end{theorem}
The proof of theorem \ref{teo2} is analogue to the proof of Theorem $4.2$ in \cite{Jungel10},
pages $38-40$:
one must consider \eqref{WHYDRO2} as Wigner equations and
use the weight functions $k(p) = \{1,\vec{p}\}$ for the first equation
in \eqref{WHYDRO2} and $k(p) = 1$ for the second equation.

Eqs. \eqref{QHE-F} are a closed system of hydrodynamic equations, indeed
$g[n_0,\vec{n},\vec{J}]$ is a function of the moments $n_0$,
$\vec{n}$, $\vec{J}$ only; however, the system \eqref{QHE-F} is
a very implicit model; in the next subsection we
will build an approximated but more explicit version of
\eqref{QHE-F} by exploiting the hypothesis we have done (semiclassical
and \eqref{LSFS}).

\subsection{First hydrodynamic model: Explicit construction.}
It is possible to write the first-order approximation of
$g[n_0,\vec{n},\vec{J}]$ given by \eqref{ghydro} by following a
strategy similar to that one employed to compute the approximation of
the equilibrium distribution for the first diffusive model. More
precisely, we consider \eqref{gbeta0}, \eqref{gbeta10},
\eqref{gbeta1v} with:
$$-a =  \bigg(\frac{|p|^2}{2} + p_k\xi_0^k + \xi_0^0\bigg)\sigma_0 +
\xi_s^0\sigma_s\,,\qquad -b = \gamma \vec{p}\cdot\vec{\sigma}\,;$$
then we impose the constraints \eqref{constraintsQHE}. Skipping the
long but straightforward computations needed, we obtain:

\begin{equation}\label{EQDH}
\begin{split}
g_0[n_0,\vec{n},\vec{J}] =& \frac{n_0}{2\pi}e^{-|\vec{p}-\vec{u}|^2/2}
+ O(\epsilon^2)\,,\\
\vec{g}[n_0,\vec{n},\vec{J}] =&
\frac{n_0}{2\pi}e^{-|\vec{p}-\vec{u}|^2/2}\left(
\frac{\vec{n}}{n_0} - \epsilon\gamma Z\right)+ O(\epsilon^2)\,,
\end{split}
\end{equation}
where:
\begin{equation}\label{om}
\begin{split}
Z =& \left(
\left(1 -
  \frac{|\vec{n}|^2}{n_0^2}\right)\frac{\vec{n}\otimes\vec{n}}{|\vec{n}|^2}
+ \omega\frac{|\vec{n}|}{n_0}\left(I - \frac{\vec{n}\otimes\vec{n}}{|\vec{n}|^2}
\right)\right) (\vec{p} - \vec{u})\\
&+\frac{\omega}{2\gamma}\left(1 - \frac{|\vec{n}|}{n_0}
\right)\frac{\vec{n}}{|\vec{n}|}\wedge[(\vec{p} -
\vec{u})\cdot\vec{\nabla}]\frac{\vec{n}}{|\vec{n}|}\,,\\
\omega =& \log^{-1}\sqrt\frac{n_0 + |\vec{n}|}{n_0 -
  |\vec{n}|}\,,\qquad
\vec{u} = \frac{\vec{J}}{n_0}\,.
\end{split}
\end{equation}

Finally, if we use Eqs.\  \eqref{EQDH}, \eqref{om} to explicitly compute the terms
in \eqref{QHE-F} up to $O(\epsilon^2)$
we find the following hydrodynamic equations:
\begin{equation}\label{QHE}
\begin{split}
\partial_t n_0 =& -\frac{\vec{\nabla}}{2\gamma}\cdot(\vec{J} +
\epsilon\gamma\vec{n}) + O(\epsilon^2)\,,\\
\partial_t\vec{n} =& -\frac{\vec{\nabla}}{2\gamma}\cdot\left(
\frac{\vec{n}\otimes\vec{J}}{n_0} - \epsilon\gamma n_0 \Phi\right) -
\frac{\epsilon}{2}\vec{\nabla}n_0\\
&-\frac{\vec{n}\wedge\vec{J}}{n_0} - \frac{\epsilon}{2}n_0\omega
\left(1 - \frac{|\vec{n}|}{n_0}\right)\left[\vec{\nabla}\cdot\left(
\frac{\vec{n}}{|\vec{n}|}\right) -
\frac{\vec{n}}{|\vec{n}|}\cdot\vec{\nabla}\right]\frac{\vec{n}}{|\vec{n}|}
+ O(\epsilon^2)\,,\\
\partial_t\vec{J} =&
-\frac{\vec{\nabla}n_0}{2\gamma}
-\frac{\vec{\nabla}}{2\gamma}\cdot\left(
\frac{\vec{J}\otimes(\vec{J}+\epsilon\gamma\vec{n})}{n_0}\right)
- n_0\vec{\nabla}V + O(\epsilon^2)\,,
\end{split}
\end{equation}
where $\omega$ is given by \eqref{om},
\begin{equation*}
\begin{aligned}
\Phi =& \left(1 -
  \frac{|\vec{n}|^2}{n_0^2}\right)\frac{\vec{n}\otimes\vec{n}}{|\vec{n}|^2}
+ \omega\frac{|\vec{n}|}{n_0}\left(I - \frac{\vec{n}\otimes\vec{n}}{|\vec{n}|^2}
\right) \\
& + \frac{\omega}{2\gamma}\left(1 - \frac{|\vec{n}|}{n_0}
\right)\left(\frac{\vec{n}}{|\vec{n}|}\wedge\right)\left[
\vec{\nabla}\otimes\left(\frac{\vec{n}}{|\vec{n}|}\right)\right]^T\,,
\end{aligned}
\end{equation*}
the superscript $T$ denotes the matrix transpose and for an arbitrary
$v\in\mathbb{R}^3$ we defined the matrix $(v\wedge)$ as the only $3\times 3$ real matrix
such that:
$$(v\wedge)z = v\wedge z\qquad z\in\mathbb{R}^3\,. $$

\section{A second hydrodynamic model for graphene}
We are going to build another hydrodynamic model for quantum electron
transport in
graphene, starting again from the Wigner equations in hydrodynamic
scaling \eqref{WHYDRO2}, considering the same fluid-dynamic moments
$n_0$, $\vec{n}$, $\vec{J}$ of the Wigner distribution $w(r,p,t)$ and
 taking again as the equilibrium distribution the one given in \eqref{ghydro},
\eqref{constraintsQHE}; however, we will make
stronger assumptions than \eqref{LSFS}, which will allow us to
consider also $O(\epsilon^2)-$terms in the fluid equations.
\subsection{Assumptions}
We perform the semiclassical approximation $\epsilon\ll 1$ and the SMS
 approximation, that is we suppose:
\begin{equation}\label{SMSH}
c = O(\epsilon)\,,\qquad|\vec{\xi}^{\,0}| = O(\epsilon)\,,
\end{equation}
where $\vec{\xi}^{\,0} = (\xi_1^0,\xi_2^0,\xi_3^0)$ are the Lagrange
multipliers appearing in \eqref{ghydro}. We define again $\gamma$ as
in \eqref{gamma}. We will see that from the
hypothesis \eqref{SMSH} will follow that:
\begin{equation}\label{SMSH2}
\frac{|\vec{n}|}{n_0} = O(\epsilon)
\end{equation}
which is the hydrodynamic analogue of the relation \eqref{SMS2} valid for the diffusive model \eqref{QDE2}.
\subsection{Second hydrodynamic model: Explicit construction}
For the sake of simplicity, let us redefine: $\vec{\xi}^{\,0}\mapsto \epsilon \vec{\xi}^{\,0}$ and consider $|\vec{\xi}^{\,0}| = O(1)$.\\
Under our hypothesis, the classical symbol of the modified Hamiltonian becomes:
\begin{equation}\label{abQHE2}
-h_\xi = a + \epsilon b\,,\quad -a =
\left(\frac{|p|^2}{2} + \xi_0^k p_k + \xi_0^0\right)\sigma_0\,,
\quad -b = (\gamma p_s + \xi_s^0)\sigma_s\,;
\end{equation}
that is, the modified Hamiltonian decouples in a \emph{scalar} part of
order $O(1)$ and a \emph{spinorial} part of order $O(\epsilon)$:
again, this fact
leads to remarkable simplifications in computations.

We can compute the second order expansion of the equilibrium
distribution \eqref{ghydro}, \eqref{constraintsQHE} under the
hypothesis \eqref{SMSH} through a
strategy similar to that one
employed to compute the second order expansion of the
equilibrium distribution for the second diffusive model:
first we substitute \eqref{abQHE2} in
\eqref{gbeta0-bis}, \eqref{gbeta1-f-bis},
\eqref{gbeta2-f}, then we impose the constraints
\eqref{constraintsQHE} (we omit the
long but straightforward computations needed). The following
result holds:
\begin{equation}\label{gqhe}
\begin{aligned}
g_0[n_0,\vec{n},\vec{J}] =& \hat{g}\left[
n_0 - n_0\left(\frac{|\vec{n}|^2}{2n_0^2} +
  \epsilon^2\gamma^2\right),\,
\vec{J} + \epsilon\gamma\vec{n} - \left(
\frac{|\vec{n}|^2}{2n_0^2} + \epsilon^2\gamma^2\right)
\vec{J}\right]\\
&+ \frac{n_0}{4\pi}e^{-|\vec{p} - \vec{J}/n_0|^2/2}
\left|\frac{\vec{n}}{n_0} - \epsilon\gamma\left(
\vec{p} - \frac{\vec{J}}{n_0}\right)\right|^2
+ O(\epsilon^3)\,,\\
\vec{g}[n_0,\vec{n},\vec{J}] =& \frac{n_0}{2\pi}e^{-|\vec{p} -
  \vec{J}/n_0|^2/2}\left(\frac{\vec{n}}{n_0} - \epsilon\gamma\left(
\vec{p} - \frac{\vec{J}}{n_0}\right)\right) + O(\epsilon^3)\,,
\end{aligned}
\end{equation}
where $\hat{g}[n_0,\vec{J}]$ is the $O(\epsilon^3)-$approximation of
the scalar quantum Maxwellian with moments:
$$\langle\hat{g}[n_0,\vec{J}]\rangle = n_0\,,\qquad \langle\vec{p}\,
\hat{g}[n_0,\vec{J}]\rangle = \vec{J}\,,$$
which can be found in \cite{JungelMatthesMilisic06}. Let us just
stress that the computations show that the SMS approximation leads to
\eqref{SMSH2}, as anticipated (see the Appendix for details).

By using the $O(\epsilon^3)$-expansion of $g[n_0,\vec{n},\vec{J}]$ to
explicitly calculate the terms inside Eqs.\  \eqref{QHE-F} up to
$O(\epsilon^3)$ we find the following hydrodynamic model:
\begin{equation}\label{QHE2}
\begin{split}
\partial_t n_0 + \frac{\vec{\nabla}}{2\gamma}\cdot(\vec{J} +
\epsilon\gamma\vec{n}) =& O(\epsilon^3)\,,\\
\partial_t\vec{n} + \frac{\vec{\nabla}}{2\gamma}\cdot\left(
\frac{\vec{n}\otimes\vec{J}}{n_0}\right)
+ \frac{\vec{n}\wedge\vec{J}}{n_0} =& O(\epsilon^3)\,,\\
\partial_t\vec{J} + \frac{\vec{\nabla}}{2\gamma}\cdot\left(
\frac{\vec{J}\otimes(\vec{J}+\epsilon\gamma\vec{n})}{n_0}\right) +
\frac{\vec{\nabla}n_0}{2\gamma}+
n_0\vec{\nabla}(V + V_B) =& O(\epsilon^3)\,,
\end{split}
\end{equation}
where $V_B$ is the Bohm potential, defined by \eqref{Bohm}.

\section{Conclusions and perspectives for the future}
In this paper we have proposed four fluid-dynamic models for quantum
electron transport in graphene: the first two ones are diffusive-type
models, the last two ones are hydrodynamic-type models. All models
have been built by using a statistical closure of the moment
equations derived from the
Wigner system \eqref{WE2} based on the minimization of the
quantum entropy functional \eqref{Aw}, \eqref{AS}.
For each class of
models (diffusive or hydrodynamic) we have have presented
two strategies of
construction: a first approach, consisting in making weaker
hypothesis (semiclassical and Low Scaled Fermi Speed) but neglecting
$O(\epsilon^2)$ terms in the fluid equations, which leads to the
models \eqref{QDE}, \eqref{QHE}; and a second approach, consisting
in making stronger hypothesis (semiclassical and Strongly Mixed State)
but neglecting $O(\epsilon^3)$ terms, which leads to the
models \eqref{QDE2}, \eqref{QHE2}.

In the next future, we are going to perform numerical simulations with
the models \eqref{QDE2} and \eqref{QHE2}, attempting to reproduce the
features of charge transport in graphene, with particular attention to
the so-called ''Klein paradox'' (unimpeded penetration of electrons
through arbitrary high potential barriers, see
\cite{Beenakker, KatsnelsonEtAl06}). Moreover we
plan to build new fluid-dynamic models for quantum transport in
graphene, again by an entropy minimization strategy, but using
Fermi-Dirac entropy instead of Maxwell-Boltzmann entropy, the former
being more suitable for describing electrons than the latter.

\appendix
\section{Proof of Eqs.~\eqref{SMS2-bis} and \eqref{SMSH2}}
From \eqref{abQDE2}, \eqref{gbeta-bis}, \eqref{gtildeQDE} follows:
\begin{equation}\label{app-1}
\begin{aligned}
g_0[n_0, n_\sigma] =& e^{-(|p|^2/2 + A)} + O(\epsilon^2)\,,\\
\vec{g}[n_0, n_\sigma] =& -\epsilon e^{-(|p|^2/2 + A)}(B + \gamma
|p|)\frac{\vec{p}}{|p|} + O(\epsilon^2)\,;
\end{aligned}
\end{equation}
from \eqref{constraintQDE2}, \eqref{app-1} we have:
\begin{equation*}
\begin{aligned}
n_0 =& \langle g_0[n_0, n_\sigma]\rangle = 2\pi e^{-A} + O(\epsilon^2)\,,\\
n_\sigma =& \langle\vec{g}[n_0, n_\sigma]\cdot\vec{p}/|p|\rangle =  -2\pi\epsilon e^{-A}\left(B +
  \gamma\sqrt\frac{\pi}{2}\,\right) + O(\epsilon^2)\,;
\end{aligned}
\end{equation*}
so \eqref{SMS2-bis} follows.

From \eqref{abQHE2}, \eqref{gbeta0-bis}, \eqref{gbeta1-f-bis},
\eqref{gbeta2-f} we obtain:
\begin{equation}\label{app-2}
\begin{aligned}
g_0[n_0,\vec{n},\vec{J}] =& e^{-|\vec{p} - \vec{\nu}|^2/2 -A} + O(\epsilon^2)\,,\\
\vec{g}[n_0,\vec{n},\vec{J}] =&-\epsilon e^{-|\vec{p} - \vec{\nu}|^2/2 -A}(
\vec{B} + \gamma\vec{p}) + O(\epsilon^2)\,,
\end{aligned}
\end{equation}
where $\vec{\nu} = -(\xi^1_0, \xi^2_0,0)$ and $A = \xi_0^0 -
|\vec{\nu}|^2/2$; so from \eqref{app-2}, \eqref{constraintsQHE} we
deduce:
\begin{equation*}
\begin{aligned}
n_0 =& \langle g_0[n_0,\vec{n},\vec{J}]\rangle = 2\pi e^{-A} + O(\epsilon^2)\,,\\
\vec{n} =&\langle\vec{g}[n_0,\vec{n},\vec{J}]\rangle=
  -2\pi\epsilon e^{-A}(\vec{B} + \gamma\vec{p}) + O(\epsilon^2)\,;
\end{aligned}
\end{equation*}
then:
$$|\vec{n}| = \epsilon n_0 |\vec{B} + \gamma\vec{p}\,| +
O(\epsilon^2)\,,$$
so \eqref{SMSH2} follows.

\section*{Acknowledgments}
The author is thankful to L.\ Barletti of the Dipartimento di
Matematica ''Ulisse Dini'', Universit\`a di Firenze, Italy, for
drawing the author's attention to the investigated problem and for
the help received from him in the research activity.
Support from INdAM-GNFM (National Group for Mathematical Physics), as well
as from Universit\`a di Firenze research funds, is acknowledged.

\end{document}